\documentclass[12pt]{article}
\usepackage{graphicx}
\usepackage[bookmarksnumbered,colorlinks,plainpages]{hyperref}

\begin{document}

\title{Nonextensive thermostatistics for heterogeneous systems containing different $q$'s}

\author{Q.A. Wang \\
Institut Sup\'erieur des Mat\'eriaux et Mécaniques Avanc\'es,
\\ 44 Avenue F.A. Bartholdi, 72000 Le Mans, France}

\date{}

\maketitle

\begin{abstract}
The nonextensive statistics based on Tsallis entropy have been so far used for the
systems composed of subsystems having same $q$.  The applicability of this
statistics to the systems with different $q$'s is still a matter of investigation.
The actual difficulty is that the class of systems to which the theory has been
applied is limited by the usual nonadditivity rule of Tsallis entropy which, in
reality, has been established for the systems having same $q$ value. In this paper,
we propose a more general nonadditivity rule for Tsallis entropy. This rule, as the
usual one for same $q$-systems, can be proved to lead uniquely to Tsallis entropy in
the context of systems containing different $q$-subsystems. A zeroth law of
thermodynamics is established between different $q$-systems on the basis of this new
nonadditivity.
\end{abstract}


\section{Introduction}
The starting point of the nonextensive statistical mechanics (NSM) of
Tsallis\cite{Tsal88} is the entropy given by
\begin{equation}                                    \label{1}
S_q=\frac{\sum_{i=1}^wp_i^q-1}{1-q}
\end{equation}
where the physical states are labelled by $i=1,2,...,w$, $q$ is a parameter
characterizing the nonextensivity of the theory, and $p_i$ is the probability for
the system to be found at state $i$. The application of the principle of maximum
entropy under appropriate constraints can lead to power law distributions
characterized by the so called $q$-exponential functionals\cite{Tsal98}. For
example, for canonical ensemble under the constraints associated with probability
normalization and average energy, we can have \cite{Tsal98,Wang01}:
\begin{equation}                                    \label{4}
p_i=\frac{1}{Z_q}[1-(1-q)\beta E_i]^{1/1-q} \;\;\; [\cdot]>0
\end{equation}
where $Z_q$ is the partition function, $E_i$ is the energy of a microstate $i$, and
$\beta$ is an intensive variable analogous to the inverse temperature in
Boltzmann-Gibbs statistics (BGS). $\beta$ is the central quantity discussed in this
work.

For a composite system $(A+B)$ whose joint probability $p_i(A+B)$ is given by the
product of the probabilities of its subsystems {\it having the same $q$}, i.e.,
$p_{ij}(A+B)=p_i(A)p_j(B)$, we have, from Eq.(\ref{1})
\begin{equation}                                    \label{2}
S_q(A+B)=S_q(A)+S_q(B)+(1-q)S_q(A)S_q(B)
\end{equation}
and, from Eq.(\ref{4})
\begin{equation}                                    \label{3}
E_{ij}(A+B)=E_i(A)+E_j(B)+(q-1)\beta E_i(A)E_j(B)
\end{equation}
Eqs.(\ref{2}) and (\ref{3}) characterize the nonadditivity of the same $q$-systems
having Tsallis entropy. In other words, they prescribe the class of nonextensive
systems to which NSM may be applied. They should be considered as the analogs (or
generalizations) of the entropy and energy additivity in BGS and play the role of
starting hypothesis specifying the limit of the validity range of NSM. It has been
proved that\cite{Santos,Abe00x} Eq.(\ref{2}) uniquely determines Tsallis entropy and
that\cite{Abe02a,Wang02a} Eqs.(\ref{2}) and (\ref{3}) are a group of necessary
conditions for the existence of thermodynamic stationarity between nonextensive
systems. When $q=1$, $S_q=S_{1}=-\sum_{i=1}^wp_i\ln p_i$, and Eqs.(\ref{2}) and
(\ref{3}) become the usual additivity assumption of entropy and energy for BGS.

The range of validity of this generalized theory is actually under study. In view of
its application to complex systems like turbulence, economic systems, fractal and
chaotic systems, a point of view \cite{Tsal03,Cohen,Letters} is more and more
accepted that NSM can be used to describe some nonequilibrium systems at stationary
state. This implies the following hypotheses:

\begin{enumerate}
\item Tsallis entropy, as an information or uncertainty measure associated with the
probability $p_i$, is valid for some nonequilibrium nonextensive systems in steady
or stationary states (time independent);

\item $p_i$ is the time invariant probability (or invariant measure\cite{Beck}) for
the system under consideration to be found at the stationary state $i$;

\item the Janyes' inference method (unbiased guess) of maximum
entropy\cite{Jaynes,Tribus} applies for these states, i.e., the time independent
probability distribution $p_i$ can be deduced from maximizing the entropy
(uncertainty) under some constraints associated with, e.g., the normalization of
$p_i$ and the averages of physical quantities determining $p_i$. In this work, $p_i$
is supposed to be uniquely determined by the energy $E_i$ of the system at
stationary state $i$.

\end{enumerate}

Due to the maximum Tsallis entropy, the framework of the conventional equilibrium
thermodynamics can be copied for stationary nonequilibrium systems. Essential for
this work is the establishment of zeroth law of thermodynamics, i.e., the intensive
variable $\beta=\frac{\partial S_q}{\partial U_q}$ (we call it {\it inverse
temperature} from now on) is uniform everywhere at stationary state, where $U_q$ is
the internal energy defined below in the section 3. It is worth noticing that, due
to the different formalisms of NSM proposed in the past 15 years, the inverse
temperature $\beta$ has been given several definitions which sometimes cause
confusion. A comment on this subject was given in ref.\cite{Wang04a}.

In what follows, we will focalize our attention on an important question about the
validity of NSM for the systems containing subsystems with different $q$'s. Up to
now, NSM has been applied only to composite systems containing subsystems with same
$q$ for which several versions of zeroth law of thermodynamics were established on
the basis of Eq.(\ref{2})\cite{Abe1,Wang02b,Wang03c}. For the systems having
different $q$'s, the attempt to apply NSM \cite{Sasaki,Abe} using Eq.(\ref{2}) is
unfruitful since, firstly, no rigorous zeroth law can be formulated, secondly, no
Tsallis entropy can be formulated as in Eq.(\ref{1}) for a compound system $A+B$
when $A$ and $B$ respectively have Tsallis entropy with their own $q_A$ and $q_B$.
This result implies that, in the actual formulation of NSM, there must be some
fundamental aspects which are incomplete or handicapped. Nature contains
nonextensive systems with different $q$ values, as shown by the applications of NSM
in the past years, and there must be also systems containing subsystems with
different $q$'s, such as a cluster of galaxies containing different galaxies with
different $q$'s, a heterogeneous materials containing elements being physically and
chemically different, a fractal containing two fractals of different dimensions and
physics, or a social group including socially (culturally, economically or
technically...) different subgroups, etc. If NSM is useful only for the systems
containing uniform distribution of $q$ value, not only its application would be
extremely limited, but worse, its general validity would be fundamentally in
danger\cite{Nauenberg} because, for different $q$-systems, no composability of
physical quantity is possible, no measurement of temperature or pressure of a
$q$-system is possible with ordinary thermometers or barometers which obey
Boltzmann-Gibbs thermostatistics with $q=1$ (or in order to measure the temperature
of a $q$-system, you have to verify that the thermometer has the same $q$ as the
system!). This is equivalent to saying that NSM is useless in view of the diversity
of physical or social systems. So this obstacle must be removed in order that NSM be
a coherent and complete theory whose formalism, just like that of BGS, should remain
the same whatever the level of the system, i.e., be hierarchically invariant.

As a matter of fact, it is often forgotten that Eq.(\ref{2}), a fundamental rule of
NSM, was only established for the subsystems with same $q$. The aforementioned
failure of the theory is in fact a failure of Eq.(\ref{2}). The aim of this paper is
to show that, by replacing Eq.(\ref{2}) with a more general nonadditivity rule for
Tsallis entropy, NSM can be applied to any system having different $q$'s. We can in
addition derive a calculus which uniquely determines the $q$ of a composite systems
from the $q$'s of its subsystems. Without loss of generality, the discussion will be
made with the $q$-exponential distribution given by Eq.(\ref{4}).

\section{A more general pseudoadditivity for Tsallis entropy}
Now let us consider $A$ and $B$, two nonextensive subsystems of a composite
nonextensive system $A+B$. It has been proved\cite{Abe02a} that the most general
pseudoadditivity (or composability) of entropy or energy prescribed by zeroth law is
the following :
\begin{equation}                                    \label{5}
H[Q(A+B)]=H[Q(A)]+H[Q(B)]+\lambda_Q H[Q(A)]H[Q(B)],
\end{equation}
which is in fact a very weak condition where $H[Q]$ is just certain differentiable
function satisfying $H[0]=0$, $\lambda_Q$ is a constant, and $Q$ is either entropy
$S$ or internal energy $U$\cite{Wang02a}. The functional form of $H(Q)$ naturally
depends on the nonadditive character of the system of interest. As shown in
\cite{Wang02a}, for a given relationship $Q=f(N)$ where $N$ is the number of
elements of the system, the finding of $H(Q)$ is trivial.

As a matter of fact, Eq.(\ref{5}) has been established\cite{Abe02a,Wang02a} for the
class of systems containing only subsystems having same $q$ value. It must be
generalized for the systems whose subsystems have different $q$'s. This
generalization is straightforward if we replace the Eq.(1) of reference
\cite{Abe02a}, i.e., $S(A+B)=f\{S(A),S(B)\}$ for uniform $q$, by
$H_{q}[S_q(A+B)]=f\{H_{q_A}[S_{q_A}(A)],H_{q_B}[S_{q_B}(B)]\}$ (or by
$H_{q}[S_q(A+B)]=H_{q_A}[S_{q_A}(A)]+H_{q_B}[S_{q_B}(B)]
+g\{H_{q_A}[S_{q_A}(A)],H_{q_B}[S_{q_B}(B)]\}$) where $H_q(S_q)$ is a functional
depending on $q$'s in the same way for the composite system as for the subsystems,
where $q$, $q_A$ and $q_B$ are the parameters of the composite system $A+B$, the
subsystems A and B, respectively. This functional $H_q(S_q)$ is necessary for the
equality to hold in view of the different $q$'s in the entropies of different
subsystems. The function $f$ (or $g$) is to be determined with the help of the
zeroth law. Now repeating the mathematical treatments described in the references
\cite{Abe02a,Wang02a}, we find
\begin{equation}                                    \label{5x}
H_q[Q(A+B)]=H_{q_A}[Q(A)]+H_{q_B}[Q(B)]+\lambda_Q H_{q_A}[Q(A)]H_{q_B}[Q(B)].
\end{equation}
Eq.(\ref{5}) turns out to be a special case of Eq.(\ref{5x}) for same $q$ systems.
In this case, if $H_q[S_q]$ is chosen to be identity function to give Eq.(\ref{2}),
then Tsallis entropy can be proved to be unique for this class of
systems\cite{Santos,Abe00x} just as the uniqueness of Shannon information measure
$S_1$ can be proved with the entropy additivity for independent subsystems having
product joint probability\cite{Shannon}. This reasoning can be extended to the
subsystems $A$ and $B$ each having its own $q$ if one chooses
$H_q(S_q)=\sum_{i=1}^wp_i^q-1=(1-q)S_q$. From Eq.(\ref{5x}), this choice implies :
\begin{eqnarray}                                    \label{2a}
(1-q)S_q(A+B) &=& (1-q_{A})S_{q_{A}}(A)+(1-q_{B})S_{q_{B}}(B) \\ \nonumber &+&
\lambda_S(1-q_{A})(1-q_{B})S_{q_{A}}(A)S_{q_{B}}(B)
\end{eqnarray}
where $q$, $q_A$ and $q_B$ are the parameters of the composite system $A+B$, the
subsystems A and B, respectively. Eq.(\ref{2a}) is not derived from Eq.(\ref{5}) but
only postulated in order to generalize Eq.(\ref{2}) according to the necessary
condition Eq.(\ref{5}) for the establishment of zeroth law. The uniqueness of
Tsallis entropy for the systems obeying Eq.(\ref{2a}) can also be proved along the
line of reference \cite{Shannon} by replacing the Axiom 3 of the information
additivity of \cite{Shannon} with Eq.(\ref{2a}) and the Eq.(19) of \cite{Abe00x},
i.e., $\left[1+(1-q)L_q(r^m)\right]=\left[1+(1-q)L_q(r)\right]^m$, by
$\left[1+(1-q_c)L_q(r^m)\right]=\prod_{l=1}^m\left[1+(1-q_l)L_{q_l}(r)\right]$,
where $L_q$ is the entropy functional, $q_c$ is the parameter of the composite
system containing $m$ subsystems each having its own parameter $q_l$ ($l=1,2,...m)$
and $r (\geq 2)$ states. The details of this proof will be given in another paper.

Now let $\lambda_S=1$, Eq.(\ref{2a}) implies :
\begin{eqnarray}                                    \label{6}
p_{ij}^q(A+B)=p_i^{q_A}(A)p_i^{q_B}(B)
\end{eqnarray}
which can be called the generalized factorization of joint probability for the
systems of different $q$'s. A possible understanding of this relationship is that
here the physical or effective probability is $p_i^q$ instead of $p_i$. This
interpretation may shed light on why it has been necessary to use $p_i^q$ or the
escort probability\cite{Tsal98} to calculated expectation. Eq.(\ref{6}) can be
written as the usual product probability $p_{ij}(A+B)=p_{i}(A)p_{j}(B)$ if and only
if $q=q_A=q_B$ for the systems having the same nonadditivity.

Eq.(\ref{6}) allows one to determine uniquely the parameter $q$ for the composite
system if $q_A$, $q_B$, $p_i(A)$ and $p_j(B)$ are given. By the normalization of the
joint probability, we obtain the following relationship :
\begin{eqnarray}                                    \label{7}
\sum_{i=1}^{w_A}p_i^{q_A/q}(A)\sum_{j=1}^{w_B}p_j^{q_B/q}(B)=1
\end{eqnarray}
which means $q_A<q<q_B$ if $q_A<q_B$ and $q=q_A=q_B$ if $q_A=q_B$. In this way, for
a composite system containing $N$ subsystems (k=1,2,...,N) having different $q_k$,
the parameter $q$ is determined by
\begin{eqnarray}                                    \label{7a}
\prod_{k=1}^N\sum_{i_k=1}^{w_k}p_{i_k}^{q_k/q}(A)=1,
\end{eqnarray}
from which we can say that $mini(q_k)<q<maxi(q_k)$.

Now if $(A+B)$ is at stationary state, according to Jaynes principle of unbiased
guess, $dS(A+B)=0$. From Eq.(\ref{2a}), we get :
\begin{eqnarray}                                    \label{8}
\frac{(1-q_{A})dS_{q_{A}}(A)}{1+(1-q_{A})S_{q_{A}}(A)}
+\frac{(1-q_{B})dS_{q_{B}}(B)}{1+(1-q_{B})S_{q_{B}}(B)}=0
\end{eqnarray}

Eq.(\ref{8}) is essentially different from
$\frac{dS_q(A)}{1+(1-q)S_q(A)}+\frac{dS_q(B)}{1+(1-q)S_q(B)}=0$ used for the
subsystems with same $q$ and will allow us to apply NSM to the systems in which the
distribution of $q$ value is not uniform.

\section{Zeroth law with nonadditive energy}
In general, for exact treatments of nonextensive systems within NSM, we have to deal
with nonadditive energy that allows the existence of thermal equilibrium and
stationarity\cite{Wang02a} and is compatible with the entropy nonadditivity.
Usually, the nonadditivity of energy is determined by the product probability.
Considering the distribution Eq.(\ref{4}) and the generalized product probability
Eq.(\ref{6}), we get :
\begin{eqnarray}                                    \label{15}
\frac{q}{1-q}\ln[1-(1-q)\beta(A+B) E_{ij}(A+B)] &=& \\
\nonumber \frac{q_A}{1-q_A}\ln[1-(1-q_{A})\beta(A) E_i(A)] \\
\nonumber + \frac{q_B}{1-q_B}\ln[1-(1-q_{B})\beta(B) E_j(B)].
\end{eqnarray}
Now let us define a deformed energy $e^{(q)}_i=\frac{1}{(1-q)\beta}\ln[1-(1-q)\beta
E_i]$ in order to write the distribution Eq.(\ref{4}) in the usual exponential form
\begin{eqnarray}                                    \label{15a}
p_i=\frac{1}{Z_q}\exp[-\beta e^{(q)}_i].
\end{eqnarray}
The introduction of this distribution into the generalized product joint probability
Eq.(\ref{6}) leads to :
\begin{eqnarray}                                    \label{16}
q\beta(A+B)e^{(q)}_{ij}(A+B)=q_A\beta(A)e^{(q)}_i(A)+q_B\beta(B)e^{(q)}_j(B).
\end{eqnarray}
Note that $e^{(q)}_i$ varies monotonically with $E_i$, so that $dE_i=0$ leads to
$de^{(q)}_i=0$ in accordance with the conservation of both $E_i$ and $e^{(q)}_i$.

In what follows, the zeroth law will be discussed within two possible formalisms of
NSM using the unnormalized expectation $U_q=\sum_ip_i^qE_i$ and the expectation
given with escort probability $U_q=\sum_ip_i^qE_i/\sum_ip_i^q$\cite{Beck},
respectively. In view of the generalized product joint probability Eq.(\ref{6}) and
the energy nonadditivity, the choice of these statistical expectations is necessary
in order to split the composite expectations of the total system into those of the
subsystems. The establishment of zeroth law would be impossible without this
splitting.

\subsection{With unnormalized $q$-expectation}

The formalism of NSM with unnormalized expectation $U_q=\sum_ip_i^qE_i$ was proposed
by Tsallis and co-workers\cite{Tsal88,Curado} and has found many
applications\cite{Tsal4,Tsal95}. Recently, it has been shown\cite{Wang03c} that,
with this expectation, the inverse temperature could be well defined by
$\beta=\frac{\partial S}{\partial U}$ for the systems of same $q$, so that all the
details of the mathematics (including the statistical interpretation of heat and
work) of BGS remain the same within NSM. In what follows, it will be shown that the
temperature can be defined between the subsystems with different $q$.

The expectation of the deformed energy is given by $u_q=\sum_ip_i^qe^{(q)}_i$. In
order to find a relationship between $\beta(A+B)$, $\beta(A)$ and $\beta(B)$, we
assume
\begin{eqnarray}                                    \label{13}
\frac{qu_q(A+B)}{\sum_{ij}p_{ij}^q(A+B)}=
\frac{q_Au_{q_A}(A)}{\sum_{i=1}p_i^{q_A}(A)}+\frac{q_Bu_{q_B}(B)}{\sum_{j=1}p_j^{q_B}(B)}
\end{eqnarray}
which is consistent with Eq.(\ref{16}) and the generalized product probability. It
is in fact the analog of the additive energy in BGS. In general, this ``additivity''
does not means $\beta(A)=\beta(B)$; it only implies
$\beta(A+B)=\frac{\beta(A)u'_{q_A}(A)+\beta(B)u'_{q_B}(B)}{u'_{q_A}(A)+u'_{q_B}(B)}$
where $u'_{q_A}=q_Au_{q_A}/\sum_{i=1}p_i^{q_A}$. From energy conservation, we obtain
$du_q(A+B)=0$ and
\begin{eqnarray}                                    \label{14}
\frac{q_Adu_{q_A}(A)}{\sum_{i=1}^wp_i^{q_A}(A)}+\frac{q_Bdu_{q_B}(B)}{\sum_{j=1}^wp_j^{q_B}(B)}=
0.
\end{eqnarray}

Now in order to determine $\beta$ in the distribution of Eq.(\ref{4}) or
Eq.(\ref{15a}), we take a deformed entropy $s_q$ defined by
Taneja\cite{Esteban,Taneja} :
\begin{equation}                                    \label{18a}
s_q=-\sum_{i=1}^wp_i^{q}\ln p_i
\end{equation}
but here $p_i$ is given by Eq.(\ref{15a}). This deformed entropy is concave and has
the following nonadditivity :
\begin{equation}                                    \label{19a}
\frac{qs_q(A+B)}{\sum_{ij}^wp_{ij}^{q}(A+B)}=
\frac{q_As_{q_A}(A)}{\sum_{i}p_i^{q_A}(A)}+\frac{q_Bs_{q_B}(B)}{\sum_{j}p_j^{q_B}(B)}
\end{equation}
which leads to, at maximum entropy,
$\frac{q_Ads_{q_A}(A)}{\sum_{i}p_i^{q_A}(A)}+\frac{q_Bds_{q_B}(B)}{\sum_{j}p_j^{q_B}(B)}=0$.
Combining this with Eq.(\ref{14}) and considering $s_q=\sum_{i}p_i^q\ln Z_q+\beta
u_q$, we get
\begin{eqnarray}                                    \label{11aa}
\beta(A)=\beta(B)
\end{eqnarray}
where the inverse temperature is given by $\beta=\frac{\partial s_q}{\partial
u_q}=\frac{\partial S_q}{\partial U_q}$. It is worth noticing that the equality of
the temperatures of $A$ and $B$ at stationary state is independent of whether $q_A$
and $q_B$ are equal.

\subsection{With escort probability}
The above description of thermal equilibrium can also be made with escort
probability proposed in \cite{Beck}. Let us define a deformed average energy with
escort probability $\mu_q=\sum_ip_i^qe^{(q)}_i/\sum_{i}p_i^q$, where $p_i$ is the
probability given by Eq.(\ref{4}). Still assuming
$q\mu_q(A+B)=q_A\mu_{q_A}(A)+q_B\mu_{q_B}(B)$ which means, from the nonadditivity
Eq.(\ref{16}), $\beta(A+B)=\frac{\beta(A)q_A \mu_{q_A}(A)+\beta(B)q_B
\mu_{q_B}(B)}{q_A\mu_{q_A}(A)+q_B\mu_{q_B}(B)}$, one gets
\begin{eqnarray}                                    \label{17}
q_Ad\mu_{q_A}(A)+q_Bd\mu_{q_B}(B)=0.
\end{eqnarray}
in accordance with the conservation of $e^{(q)}_i$ or $\mu_q$.

Now let us take the entropy $\psi_q$ defined by Aczel and
Daroczy\cite{Esteban,Aczel} as a deformation of the nonextensive entropy $S_q$ :
\begin{equation}                                    \label{18}
\psi_q=-\frac{1}{\sum_{i=1}^wp_i^{q}}\sum_{i=1}^wp_i^{q}\ln p_i
\end{equation}
which naturally arises from the escort probability. $\psi_q$ can be maximized with
appropriate constraints associated with the deformed average energy $\mu_q$ to give
the distribution of Eq.(\ref{15a}).

The above deformed entropy has the following nonadditivity :
\begin{equation}                                    \label{19}
q\psi_q(A+B) = q_{A}\psi_{q_A}(A)+q_{B}\psi_{q_B}(B),
\end{equation}
so that at maximum entropy $d\psi_q(A+B)=0$, we have
$q_{A}d\psi_{q_A}(A)+q_{B}d\psi_{q_B}(B)=0$. Combining this with equation
Eq.(\ref{17}), one gets
\begin{eqnarray}                                    \label{20}
\frac{\partial \psi_{q_A}(A)}{\partial \mu_{q_A}(A)}=\frac{\partial
\psi_{q_B}(B)}{\partial \mu_{q_B}(B)}.
\end{eqnarray}
On the other hand, from Eqs.(\ref{15a}) and (\ref{18}), it is easy to write
\begin{equation}                                    \label{21}
\psi_q=\ln Z_q+\beta \mu_q
\end{equation}
which means $\beta=\frac{\partial \psi_q}{\partial \mu_q}$. We finally have
\begin{equation}                                    \label{22}
\beta(A)=\beta(B)
\end{equation}
which is also independent of whether $q_A$ and $q_B$ are equal.

\section{An alternative method}
In the above discussions, the deformed entropies $s_q$ and $\psi_q$ are introduced
in order to discuss nonextensive statistics with the language of extensive
statistics BGS. $s_q$ and $\psi_q$ are in fact alternative expressions of $S_q$
associated with the deformed energy $u_q$ and $\mu_q$, respectively. They can be
maximized, just as $S_q$, to give $q$-exponential distribution\cite{Wang02bb} in
terms of the deformed energy $e^{(q)}_i$ [Eq.(\ref{15a})]. The description of the
stationarity characterized by Eq.(\ref{11aa}) or Eq.(\ref{22}) is of course
independent of their introduction. In what follows, I we briefly present an
alternative method without these deformations.

Using the product joint probability Eq.(\ref{6}) and the relationship
$\sum_ip_i^q=Z_q^{1-q}+(1-q)\beta U_q$ calculated from the distribution Eq.(\ref{4})
and the unnormalized expectation $U_q$, we get
\begin{eqnarray}                                    \label{xx10}
& Z_q^{1-q}(A+B)+(1-q)\beta(A+B) U_q(A+B) \\\nonumber & =
[Z_{q_A}^{1-q_A}(A)+(1-q_A)\beta(A) U_{q_A}(A)][Z_{q_B}^{1-q_B}(B)+(1-q_B)\beta(B)
U_{q_B}(B)].
\end{eqnarray}
Considering the total energy conservation $dU_q(A+B)=0$, we obtain
\begin{eqnarray}                                    \label{xx11}
\frac{(1-q_{A})\beta(A)dU_{q_A}(A)}{\sum_ip_i^{q_A}(A)}
+\frac{(1-q_{B})\beta(B)dU_{q_B}(B)}{\sum_ip_i^{q_B}(B)}=0
\end{eqnarray}
which suggests following energy nonadditivity
\begin{eqnarray}                                    \label{xx12}
\frac{(1-q_{A})dU_{q_A}(A)}{\sum_ip_i^{q_A}(A)}
+\frac{(1-q_{B})dU_{q_B}(B)}{\sum_ip_i^{q_B}(B)}=0
\end{eqnarray}
as the analog of the additive energy $dU(A)+dU(B)=0$ of BGS. From Eq.(\ref{xx12})
and Eq.(\ref{8}) follows $\beta(A)=\beta(B)$ with $\beta(A)=\frac{\partial
S_{q_A}(A)}{\partial U_{q_A}(A)}$ and $\beta(B)=\frac{\partial S_{q_B}(B)}{\partial
U_{q_B}(B)}$.

\section{Conclusion}
We have discussed the establishment of the zeroth law for stationary state between
nonextensive systems with different $q$'s. According to our starting hypotheses,
this stationary state maximizes Tsallis entropy so that the zeroth law of the
equilibrium thermodynamics can apply. This work is carried out with a generalized
nonadditivity rule of Tsallis entropy which reduces to the usual one if all the
subsystems have the same $q$. It should be noticed that the generalized product
joint probability Eq.(\ref{6}) derived from the generalized nonadditivity of entropy
suggests two possible expectation calculus of NSM, the unnormalized $q$-expectation
and the escort probability $q$-expectation, in order to establish zeroth law. A main
result of this work is to show that the intensive variable $\beta$ is uniform in the
nonextensive systems in stationary state, independently of whether or not the
systems contain subsystems with different $q$ values. So NSM theory based on the
maximization of Tsallis entropy is, contrary to some belief, also valid for the
systems containing non uniform distribution of $q$ values.

\end{document}